\documentstyle[eqsecnum,floats,prd,aps]{revtex}
\begin{document}
\draft
\preprint{UTPT-95-03}
\title{The Unholey Solution to Black Hole Information Loss}
\author{ Neil J. Cornish}
\address{Department of Physics, University of Toronto \\ Toronto,
Ontario M5S 1A7, Canada}
\twocolumn[
\maketitle
\widetext
\begin{abstract}
The simplest solution to the black hole information loss problem is
to eliminate black holes. Modifications of Einstein gravity which
accomplish this are discussed and the possibility that string theory
is free of black holes is considered.
\end{abstract}
]
\narrowtext

\section*{Introduction}
In recent years, the application of quantum mechanics to black holes has
been considered the best arena in which to find clues to theoretical physics'
holy grail -- a quantum theory of gravity. The possible connection between
gravity, quantum mechanics and thermodynamics embodied in the laws of
black hole thermodynamics is seen as an important step on the road to a
deeper understanding of quantum gravity.
The seductive beauty of these results is in itself enough to make any
suggestion
that black holes should be eliminated a highly unpalatable thesis. However,
beauty can sometimes be beguiling. We believe it can be argued that black hole
thermodynamics, and the information loss problem, are merely symptoms of the
unphysical causal division of spacetime engendered by black holes. From this
standpoint, black holes are seen as a signal that the classical gravity theory
which predicts them has broken down and a replacement should be sought.

 By definition, a black hole divides a spacetime $(M,g_{\mu \nu})$ into
two causally disconnected regions such that $M=B+J^{-}({\cal I}^{+})$, where
$B$
is the black hole region and $J^{-}({\cal I}^{+})$
is the causal past of future null infinity. This causal disconnection is the
root cause of the information loss problem in spacetimes where a black hole
is produced by the collapse of matter and subsequently evaporates away via
the Hawking process.

The three main approaches to solving the information loss problem can be
roughly classified as 1) reconciling quantum mechanics with non-unitary
evolution\cite{hawk,ellis,bsp,unruh} 2) reformulating gravity to store or
return the information\cite{aharon,banks,suss,thooft}
3) there is no problem\cite{page,bekenstein}. Each of these
possibilities is supported by ingenious and plausible arguments, and one
of the three might well be correct. However, the question of which approach is
correct is rendered moot if we insist that black holes do not exist.

The motivation for eliminating black holes goes beyond achieving a quick
fix to the information loss problem. Additional motivation is provided by
considering the Hawking-Penrose singularity theorems and the role that
trapped surfaces play in forcing singular behaviour. Moreover, black
hole event horizons cause physical measurables such as redshifts to diverge.
In the spirit of the cosmic censor conjecture and the chronology protection
conjecture of general relativity we shall demand that the entire spacetime
manifold lies in the causal past of future null infinity, i.e.
$M=J^{-}({\cal I}^{+})$. This global causality demand also requires that the
spacetime is free of singularities in order for the spacetime to be
strongly asymptotically predicatable, as we have sacked the cosmic censor.

Clearly, Einstein gravity is at odds with the no black hole condition and
alternative theories must be sought which ensure causal connectivity. In
section I we shall consider what form the modifications to Einstein's theory
must take to eliminate black holes. A concrete example which appears to satisfy
the global causality demand is reviewed in section II. The possibility that
string theory may be compatible with the no-black hole conjecture is considered
in section III.

\section{Modifying Gravity}
In physical terms, the elimination of black holes comes down to
ensuring that gravity never gets so strong, or spacetime so bent, that
light cannot escape from regions of spacetime. This probably cannot be
achieved in any theory which employs point particles, obeys the weak
equivalence principle
and is entirely local.

What we are seeking is essentially a redshift-limited theory in which the
redshift between any two points in spacetime is finite. The difficulty is that
redshift is an intrinsically non-local quantity, and any attempt to
construct a theory based on non-local notions is likely to produce acausal
effects far worse than the problems of causal disconnection it is trying to
solve. More promising possibilities are offered by theories which violate
the weak equivalence principle or employ extended structures such as strings.

In a theory which violates the weak equivalence principle local measurements
can be made to determine the strength of the gravitational field. A freely
falling observer would be able to tell that a high redshift surface was being
approached. The gravitational field can respond to such
information. One might try and formulate such a theory along the lines of
the Limited Curvature Hypothesis\cite{brand} by explicitly constructing a
limited redshift Lagarangian. In section II we shall review a different
approach in which an equivalence principle violating theory, formulated
on a non-Riemannian manifold, is able to eliminate black holes from static,
spherically symmetric spacetimes.

In a theory which employs extended objects such as strings, there is an
essential non-locality built into the physics which allows redshifts to be
felt. However, since the non-locality
is expected to be confined to Plankian scales, it may seem impossible for
string theory to have any impact on macroscopic horizons. This is not the
case\cite{suss}.
Consider a fully stringy geometry near a high redshift surface. As
quantum fluctuations on this geometry propagate towards the surface of high
redshift they become amplified, leading to a large back reaction in the
underlying geometry. In this picture, a would-be event horizon must be
described by non-perturbative string theory. In section III we shall develop
arguments which suggest that string theory might be free of black hole
regions.

\section{Non-Riemannian Gravity}
By formulating gravity on a hypercomplex non-Riemannian manifold, the
gravitational theory is endowed with additional degrees of freedom. In
particular, the metric is no longer symmetric. Coordinates can be chosen
such that the symmetric part of the metric is locally Minkowskian, however
the skew components of the metric cannot be chosen to vanish.

Recently it was shown that one such theory was free of black holes and
curvature singularities for static, spherically symmetric spacetimes\cite{us}.
The metric is given by
\begin{eqnarray}
&&ds^2=\gamma(r)dt^2-\alpha(r) dr^2-r^2 d\theta^2-r^2 \sin\theta^2 d\phi^2,
\nonumber \\
&& \hspace{0.8in}
g_{[\theta \phi]}=f(r) \sin\theta \; .
\end{eqnarray}
The coordinates have been chosen so that circles of radius $r$ have
circumference $2\pi r$. The hypercomplex skew field $g_{[\mu \nu]}$
looks superficially like the Kalb-Ramond axion of low energy string theory,
but it has significantly different dynamics. Indeed, for large $r$ we find
that $g_{[\theta \phi]}=Q_{{\rm top}} \sin\theta$ which corresponds to a
purely topological contribution with topological charge $Q_{{\rm top}}$
in four dimensional string effective theories.

The constant $Q_{{\rm top}}$ is found to equal $sM^2 /3$ where $M$ is the
ADM mass and $s$ is a dimensionless constant. Since the skew contribution
to the gravitational field does not vanish in a freely falling frame, it
enables local measurements to be made of the redshift. If we define
$\nu=-2\ln(z+1)$, where $z$ is the redshift between $r$ and spatial infinity,
we find that the invariant $F^2=g_{[\mu\nu]}g^{[\mu\nu]}=f^2/(r^4+f^2) $ can be
written as
\begin{equation}
F={|\sinh(a\nu)\sin(b\nu)+s(1-\cosh(a\nu)\cos(b\nu)|\sqrt{1+s^2}\over
|\cosh(a\nu)-\cos(b\nu)|} \;
\end{equation}
where
\newpage

\
\begin{figure}[h]
\vspace{47mm}
\includegraphics{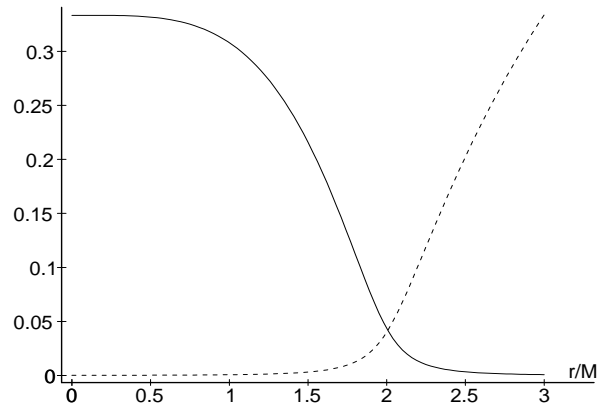}
\vspace{5mm}
\caption{The skew invariant $F^2/3$ (solid line) and the metric component
$\gamma$ (dashed line) for the choice $s=1$ in the high redshift region.}
\end{figure}
\begin{equation}
a=\sqrt{{\sqrt{1+s^2}+1 \over 2}}\; , \hspace{0.3in}
b=\sqrt{{\sqrt{1+s^2}-1 \over 2}}\; .
\end{equation}
The above equation provides an implicit expression for the redshift in terms
of a locally measurable quantity, $F$. When $F<<1$ we can recast the theory
as Einstein gravity non-minimally and non-polynomially coupled to a skew field.
 From this standpoint we can study the linearised equation of motion for $f$ in
a Schwarzshild background. Near $r=2M$ we find that
\begin{equation}
 f''\approx { 1\over M(r-2M)}\left(-Mf'
 +2f \right) \; . \label{amp}
\end{equation}
Outside of $r=2M$, $f$ can be very small and slowly decreasing $(f,f' <\! <
0)$.
Near $r=2M$, $f''$ becomes very large which implies $-f'$ becomes large
also. This means that $f$ must be rising precipitously as $r=2M$ is
approached from outside. This is confirmed by the exact solution which
is displayed graphically in Fig.1. We see that the skew field responds to
the increasing redshift by becoming very large. In turn, the back-reaction of
the skew field on the symmetric metric ensures that the redshift,
$z_{12}=\gamma_{1}^{-1/2}-\gamma_{2}^{-1/2}$,
remains finite everywhere. Importantly, the two non-vanishing curvature
invariants, $R_{\mu\nu\kappa\lambda}R^{\mu\nu\kappa\lambda}$ and
$R_{\mu\nu\kappa\lambda}R^{\rho\delta\kappa\lambda}
R_{\rho\delta}^{\;\; \mu\nu}$, are everywhere finite also.
The region below $r=2M$ is characterised by having very large, but finite
redshifts relative to spatial infinity. In this respect the solution describes
a grey, rather than a black hole. In a practical sense it is difficult to
transmit information from below $r=2M$, but there is no barrier in principle.

An important feature of the solution is that Einstein gravity can only be
recovered as a limit to this theory when the redshift is small.
In the limit $s \rightarrow 0$ the theory is identical to Einstein's theory
outside of $r=2M$, but no matter how small the parameter $s$ is taken to be,
Einstein gravity cannot be recovered at or below $r=2M$\cite{me,sok}. In
the small $s$ limit the solution has the following interesting features:

\noindent 1) The maximum redshift between any two points in the manifold is
given by
\begin{equation}
z_{{\rm max}}\approx \exp\left({\pi \over 2 |s|}\right) \; ;
\end{equation}
2) The maximum curvature occurs at $r=0$ and is given by
\begin{eqnarray}
&& R_{\mu\nu\kappa\lambda}R^{\mu\nu\kappa\lambda} \approx
{1 \over M^4}\exp\left({2\pi+4|s|\ln s+4s-9|s|\ln 2 \over |s|}\right)
\nonumber \\ \nonumber \\
&& \hspace{0.78in} \approx \left({z_{{\rm max}} \over M}\right)^4\; ;
\end{eqnarray}
3) The proper surface area at $r=2M$ and $r=0$ approaches $V_{{\rm S}}=16\pi
M^2$ as $s\rightarrow 0$:
\begin{eqnarray}
&& V_{2M}\approx V_{{\rm S}}\left(1+\frac{3}{16}s^2(\ln |s|)^2\right)\; , \\
&& V_{0}\approx V_{{\rm S}}\left(1-{|s|\pi \over 8}\right)\; ;
\end{eqnarray}
4) The proper distance between $r=0$ and $r=2M$ shrinks to zero as $s$
tends to zero since
\begin{equation}
\int_{0}^{2M}\sqrt{\alpha} \, dr \approx {4M \over z_{2M}} \; ,
\end{equation}
where $z_{2M}$ is the redshift between $r=2M$ and $r=\infty$.

When $s=0$ the redshifts $z_{2M}$ and $z_{{\rm max}}$ become infinite and the
proper distance between $r=0$ and $r=2M$ goes to zero. The curvature at the
horizon diverges when $s=0$, giving rise to a light-like singularity at $r=2M$.
Unlike Einstein's theory, this redshift dependent theory considers horizons
on the same footing as curvature singularities - for any non-zero value of $s$
it gets rid of both at the same time.

Similar results continue to hold for the analog of electrically charged
black holes, and there are promising signs that rotation will not alter
our conclusions. In summary, the theory appears to be both free of black holes
and asymptotically predictable. A general proof for arbitrary symmetry
remains to be found.

\section{String Gravity}
It is commonly hoped that string theory will temper curvature singularities and
provide a consistent theory of quantum gravity. While that might seem a lot
to ask, we want even more - we ask that string theory also banishes black
holes.

The standard picture of string theory, or at least string effective
theory, holds that the string corrections to general relativity should be
important when curvatures approach the string scale $\alpha'$.
 From this standpoint it would seem nonsensical to suggest that string theory
should offer any assistance in removing horizons, as curvatures can be
arbitrarily small at a horizon. We do not believe the possibility is that
easily dismissed.

In order to be consistent with the no-black hole condition and asymptotic
predictability, a theory must be free of singularities. We will assume that
string theory lives up to its promise and provides us with non-singular
solutions. We hope to motivate the possibility that string theory also removes
horizons by appealing to three main arguments. These arguments are based on
1) string effects in black hole backgrounds; 2) the nature
of lowest order string black hole solutions and duality transformations; and
3) the unitarity of full string theory.

The first suggestion that string theory should have something to say about
horizons is provided by studies into the behaviour of strings in fixed
black hole backgrounds. These studies indicate that strings are not only
important\cite{suss} in the description of physics near a high redshift
surface, but that string perturbation theory breaks down in such
regions\cite{bar,strom}. The fixed background perspective in these studies
requires that a complementarity principle is invoked to reconcile the
non-perturbative stringy effects seen by static observers and the total
absence of stringy effects seen by free fall observers. However, if we consider
the fact that the background should also be described by string theory we
arrive
at a somewhat different picture. In many respects the background geometry
shares the viewpoint of a static observer.
The high redshift surface suggested by the lowest order solution
serves to excite the string modes, requiring a higher order, non-perturbative
description. Via this mechanism, which is in many respects similar to the
field amplification seen in (\ref{amp}), it may be possible for the full
solution to mollify the infinite redshifts suggested by the lowest order
solution.

The second suggestion comes from considering solutions to the lowest order
metric-dilaton string Lagrangian\cite{gsw}
\begin{equation}
{\cal L}={1 \over 8\pi \alpha'}\sqrt{-g}e^{\phi}\left(R(g)
+(\nabla \phi)^2\right) \; .
\end{equation}
In four dimensions the spherically symmetric solution describes
a black hole with metric\cite{burg}
\begin{eqnarray}
&&ds^2=\left(1-{2M \over r\cos\psi}\right)^{\cos\psi}dt^2
-\left(1-{2M \over r\cos\psi}\right)^{-\cos\psi}dr^2 \nonumber \\
&&\hspace{0.4in}-r^2\left(1-{2M \over r\cos\psi}\right)^{1-\cos\psi}
d\Omega^2 \; ,
\end{eqnarray}
and dilaton
\begin{equation}
e^{\phi}=\left(1-{2M\over r\cos\psi}\right)^{\sin\psi} \; .
\end{equation}
The constant $M$ is the ADM mass of the black hole and $\psi$ is a constant
which can take any value.
For all configurations with a non-trivial dilaton $(\sin\psi \neq 0)$ both
the origin at $r=0$ and the horizon at $r=2M/\cos\psi $ suffer from
infinite curvatures, rendering this lowest order solution invalid on both
surfaces. An interesting feature of the solution is that it is
invariant under the transformation
\begin{equation}
r=r'+{2M \over \cos\psi}\; ,\quad \psi = \psi ' +\pi \; ,
\end{equation}
which interchanges the horizon and the singularity. This is reminiscent of the
duality transformation which interchanges the singularity and horizon in
(1+1)-dimensional black hole solutions to string theory\cite{wit,verl}. In
the (1+1)-dimensional case a regular horizon gets mapped into a singularity
and vice-versa. Both these classical black hole solutions to string theory
suggest that if string theory is going to help remove singularities, it should
also have something to say about horizons.

An alternative viewpoint might be that a singularity in the manifold associated
with $g_{\mu\nu}$ does not necessarily correspond to what constitutes a true
singularity in string theory as strings ``feel'' a richer geometry than point
particles. In other words, the stringy spacetime foam cannot be effectively
modeled by a smooth manifold which is free of curvature singularities.
In that event, the preceding argument fails. However, the notion
of a black hole region in the manifold associated with $g_{\mu\nu}$ would not,
by the same reasoning, necessarily correspond to a black hole region as seen by
strings. In this way the no-black hole condition might be satisfied in a
subtler sense.

Support for the notion that strings are consistent with the no-black hole
hypothesis comes from the observation that full string theory is unitary, i.e.
outgoing and incoming density matrices are related by
\begin{equation}
\rho_{{\rm out}}= S\hspace*{-6pt} / \;\rho_{{\rm in}} \; ,
\end{equation}
where $S\hspace*{-5.5pt}/\hspace*{3pt} $ is factorizable as the product
$S\, S^{\dag}$.
In standard point particle field theory, or light particle string
field theory, the presence of black hole regions in the background spacetime
destroys the factorizability of $S\hspace*{-5.5pt}/\hspace*{3pt}$. This is an
alternative way of describing the black hole information loss problem.

One approach to resolving the discrepancy between the behaviour of the full
theory and its approximate low energy description is to formulate self
consistent string modifications of quantum mechanics which allow for apparent
non-unitary evolution in the effective light particle string
theory\cite{ellis}. While this is a valid approach, it can be viewed as being
somewhat roundabout: a well behaved exact theory is replaced by
a low energy approximation with black holes; the black holes in turn cause
the low energy description to violate unitarity; the loss of unitarity requires
quantum mechanics to be modified; the non-unitary evolution cancels out
the causal disconnection so that we finally recover the consistent
description we had at the outset.
A simpler reconciliation is offered by the no-black hole hypothesis where
we have argued that fully stringy background spacetimes are free of black
holes. In this scenario both light particle string field theory and
full string theory are described by unitary evolution, allowing a smooth
recovery of standard quantum field theory in the point limit.

\section{Conclusions}
We have advanced the view that black holes are physical pathologies which
signal the breakdown of any classical gravity theory which predicts they
exist. Our viewpoint is at least consistent with observations, as there is no
direct evidence that black holes exist. By exhibiting one concrete
example, based on an equivalence-principle-violating modification of Einstein's
theory, we have demonstrated that the no-black hole hypothesis is not
an impossible demand. By putting forward heuristic arguments, we hope
to have made plausible our suggestion that string theory also banishes
black holes. At the very least, we see no reason why full string theory must
have black holes.

\section*{Acknowledgments}
I am grateful for the support provided by a Commonwealth
Scholarship. I thank Janna Levin for discussions and sharing
related ideas on low energy string theory with me.

\end{document}